\documentclass[aps,prd,twocolumn]{revtex4-1}
\usepackage{tabularx}
\usepackage{graphicx}
\usepackage{amsmath}
\usepackage{amssymb}
\usepackage{comment}
\usepackage{color}
\usepackage[svgnames]{xcolor}
\usepackage[colorinlistoftodos]{todonotes}

\newcommand{\nruns}{N_{\rm runs}}
\newcommand{\niter}{N_{\rm iter}}

\begin{document}
\title{Particle Swarm Optimization based search for gravitational waves from compact binary coalescences: performance improvements}
\author{Marc E.~Normandin}
\affiliation{Dept. of Physics and Astronomy, University of Texas San Antonio, One UTSA Circle, San Antonio, TX 78249}
\author{Soumya D.~Mohanty}
\affiliation{Dept. of Physics and Astronomy, University of Texas Rio Grande Valley, One West University Blvd.,
Brownsville, Texas 78520}
\author{Thilina S.~Weerathunga}
\affiliation{Data One Global, 2701 Dallas Parkway, Plano, TX 75093}
\date{5 June 2018}

\begin{abstract}

While  a fully-coherent all-sky search is known to be optimal for detecting  signals from compact binary coalescences (CBCs), its high computational cost has limited current searches to less sensitive coincidence-based schemes. For a network of first generation GW detectors, it has been demonstrated  that Particle Swarm Optimization (PSO) can reduce the computational cost of this search, in terms of  the number of likelihood evaluations, by a factor of  $\approx 10$ compared to a grid-based optimizer. Here, we extend the PSO-based search to a network of second generation detectors and present further substantial improvements in its performance by adopting the local-best variant of PSO and an effective strategy for tuning its configuration parameters. It is shown that a PSO-based search is viable over the entire binary mass range relevant to second generation detectors at realistic signal strengths. 

\end{abstract}
\maketitle

\section{Introduction} 
\label{intro}

The 
advanced Laser Interferometric Gravitational-wave Observatory
(LIGO)~\cite{2016PhRvL.116m1103A} and the Advanced Virgo~\cite{0264-9381-32-2-024001}
detectors have detected gravitational waves (GWs) from 
several compact binary coalescence (CBC) events over their recently concluded observation runs.
Among the detected signals, GW150914~\cite{PhysRevLett.116.061102}, GW151226~\cite{PhysRevLett.116.241103}, and 
GW170104~\cite{PhysRevLett.118.221101} are binary black hole (BBH) mergers
detected in two-way coincidence between the LIGO detectors, while 
GW170814~\cite{abbott2017gw170814} is a BBH merger detected in three-way coincidence
after Virgo joined the LIGO observation runs. 
GW170817~\cite{PhysRevLett.119.161101}, the final event in the last observing run, was a 
binary neutron star (BNS) 
inspiral. Its prompt localization on the sky by the LIGO-Virgo 
network allowed the detection of an electromagnetic (EM) counterpart~\cite{GW170817GWEM},
establishing BNS mergers as the source of some short gamma-ray bursts.

Over the next few years, LIGO and Virgo will be joined by the KAGRA~\cite{2012CQGra..29l4007S} detector in Japan and LIGO-India~\cite{2013IJMPD..2241010U}. 
Combining the data from this network of geographically 
distributed second generation 
detectors will significantly increase both the detection sensitivity
and sky localization of CBC sources~\cite{fairhurst2009triangulation,0004-637X-739-2-99}.

It is known that the 
optimal data analysis methods for the detection and estimation of  
CBC signals with a network of GW detectors are the
Generalized Likelihood Ratio Test (GLRT) and Maximum Likelihood Estimation (MLE) respectively~\cite{kay_vol1}.
In the GW literature, the GLRT is called the fully coherent all-sky search. (For a
single detector, the GLRT is the matched filter~\cite{Helstrom}.)

In both GLRT and MLE, the joint likelihood function of data from a network
of detectors needs to be maximized over the space of CBC signal parameters. However, the computational cost of
this task 
has proven to be a limiting factor in running the GLRT  as an always-on search.
The difficulty of the optimization stems from a combination of
(i) the high dimensionality of the 
search space,  and (ii) the high computational cost of evaluating the likelihood at a
given point in the search space. The latter is due to the requirement of correlating
pairs of time series that involve $O(10^6)$ samples each (for low binary component masses). 

To circumvent the computational bottleneck above, current CBC searches use a coincidence-based semi-coherent approach in which
 the GLRT is only used when (i) computationally cheaper single-detector matched filter searches result in the crossing of 
preset detection thresholds in any pair of detectors, and (ii) 
the corresponding estimated signal parameters are proximal within some preset tolerances.
As shown in~\cite{Macleod_2016}, a semi-coherent search 
trades-off a significant amount of sensitivity for the 
reduced computational cost, with the detection volume being $\sim 25\%$ smaller
than a fully-coherent search.

It has recently been demonstrated in~\citep{2017PhRvD..95l4030W}, henceforth referred to as WM, that 
Particle Swarm Optimization (PSO)~\citep{Kennedy_95}, a family of 
stochastic global optimization methods based on the behavior of biological swarms, 
can potentially solve
this challenge. Simulation of a four detector network, coinciding with the location and orientation of
the LIGO Hanford (H), LIGO Livingston (L), Virgo (V), and KAGRA (K), showed that the 
number of likelihood evaluations required by PSO is $\sim 1/10$ that of a grid-based search 
in which the search space is populated densely with a fixed set of sampling locations. 

The results in WM  were derived under the following
limitations. (i) It was 
 assumed that the noise in each detector had a Power Spectral Density (PSD) given by the 
initial LIGO design sensitivity curve~\cite{Lazzarini_96}. 
This was primarily done to keep signal lengths short ($\lesssim 30$~sec) and the computational run time manageable given the \textsc{matlab}~\cite{matlab} based
implementation of the search. (ii) As with any stochastic optimization
algorithm, PSO is not guaranteed to converge to the global optimum. For 
the so-called global-best variant of PSO used in WM,
this led to a small but non-zero reduction in detection 
probability of $\lesssim 2.5\%$. (iii) A PSO-based search
can be parallelized efficiently 
in a multi-processor environment but the code did not implement all of 
the possible layers of parallelization. 

In this paper, we develop the PSO-based fully-coherent all-sky search 
further by overcoming the above limitations. (i) We shift to the 
advanced LIGO design sensitivity and examine the performance of PSO
for both short ($O(1)$~min) and long ($O(30)$~min) data lengths.
(ii) We use the local-best variant of PSO and
find it to 
be significantly better in terms of convergence, achieving $\lesssim 0.5\%$ loss in
detection probability for $\approx 1$~min long data 
at lower signal strengths than the one used in WM. (iii) Besides translating the code into the C language, a two-layered parallelization scheme is implemented that speeds up execution by 
a factor of $\approx 7.5$.

The rest of the paper is organized as follows.
Sec~\ref{network} sets up the data model used in this paper and
provides pertinent details of the
 fully-coherent all-sky search. 
Sec.~\ref{pso} describes the local best 
PSO algorithm and a general purpose strategy for tuning its performance
on parametric GW data analysis problems.
The simulation setup and
results on detection and estimation performance of the PSO-based search are presented in Sec.~\ref{Results}. The computation time of the 
current code is discussed in Sec.~\ref{sec:latency} along with possible future
avenues that could reduce it substantially. The conclusions from our 
study are presented in Sec ~\ref{Conclusion}.


\section{Fully-coherent all-sky search}
\label{network}
While the review of the data model and the fully-coherent all-sky search 
 in this section provides a self-contained background
 for the rest of the paper, we refer the reader to WM and references therein
for a comprehensive presentation of 
mathematical details.

\subsection{Data and Signal Models}
\label{datamodel}
Consider  the ${i}^{\rm th}$ detector in a network of $D$ detectors.  
A segment of the strain time series recorded by the detector is given by
\begin{equation} 
\begin{aligned}
x^{i}(t) = \left\{\begin{array}{lc}n^{i}(t);& H_0,\\
                                   h^{i}(t) + n^{i}(t);& H_1
                                   \end{array}
                                   \right.
 \\
\end{aligned}
\end{equation}
where $h^i(t)$ is the detector response to the incident GW and $n^i(t)$ denotes detector
noise. $H_0$ and $H_1$ correspond to the two hypotheses one can propose 
about the data where
a signal is, respectively, absent or present. 

We will assume that $n^{i}(t)$ is a realization of a zero-mean, stationary Gaussian
stochastic process,
\begin{eqnarray}
E[n^{i}(t)] &=& 0;\\ 
E[n^{i}(t)n^{i}(t^\prime)] & = & \frac{1}{2}\int_{-\infty}^\infty df e^{2\pi j f (t-t^\prime)} S_n(f)\;,
\end{eqnarray}
with $S_n(f)$ denoting the one-sided noise power spectral density (PSD). It does not carry a detector
index in this paper because we assume identical PSD for all the detectors.

For a source located at azimuthal angle $\alpha$ and  polar angle $\delta$ in the Earth Centered Earth Fixed Frame (ECEF)~\cite{Leick_04}, the detector responses are given by,
\begin{eqnarray}
\left( \begin{array}{c}
h^1(t+\Delta^1(\alpha,\delta))\\
h^2(t+\Delta^2(\alpha,\delta))\\
\vdots\\
h^D(t+\Delta^D(\alpha,\delta))
\end{array}
\right) & = & {\bf F}(\alpha,\delta,\psi)\left(
\begin{array}{c}
h_+(t)\\
h_\times(t)
\end{array}
\right)\;,
\label{eq:detresponse}
\end{eqnarray}
where the ${i}^{\rm th}$ row of the antenna pattern matrix ${\bf F}(\alpha,\delta,\psi)$ contains the antenna pattern functions
$ (F_{+}^i(\alpha,\delta,\psi),\; F_{\times}^i (\alpha, \delta,\psi))$
of the ${i}^{\rm th}$ detector, $h_+(t)$ and $h_\times(t)$ are the TT gauge polarization components
of the GW plane wave 
incident on the origin of the ECEF, and $\Delta^i(\alpha,\delta)$ is the time delay between the plane wave hitting the ECEF origin and the ${i}^{\rm th}$ detector. The polarization angle $\psi$ gives the orientation of the wave frame axes with respect to 
the fiducial basis formed by  $-\widehat{\alpha}$ and $\widehat{\delta}$ in the plane orthogonal to the wave propagation direction. 

 The condition number~\cite{2006CQGra..23S.673R} 
 of the antenna pattern matrix as a function of $\alpha$ and $\delta$ 
plays an important role in determining the errors in the estimation of the source location~\cite{PhysRevLett.118.151104}. In general, the errors
worsen as the condition number at the source location increases. 
However, as will be apparent in Sec.~\ref{sec:estimation}, the influence of the condition number is more subtle than this empirical rule of thumb.

In this paper, we use a four-detector network consisting of the two LIGO detectors at Hanford (H) 
 and Livingston (L), Virgo (V) and KAGRA (K). For simplicity, we
 assume the  advanced LIGO design PSD~\cite{aligopsd}
 corresponding to the ``ZERO DET high P'' configuration for the noise in each detector. 
The orientations and locations
 of the detectors, listed in table II of WM,  match their real-world values.
 A sky map of the condition number of the antenna pattern matrix for the above detector network is shown in Figure 4 of WM. 

The polarizations $h_+(t)$ and $h_\times(t)$ used in this paper are given by
the restricted 2-PN waveform 
from a circularized binary consisting of non-spinning compact objects~\cite{Blanchet_95} that depends on
the following parameters. (i) The component masses $m_1$ and $m_2$ 
expressed through the chirp times 
$\tau_0$ and $\tau_{1.5}$,
\begin{eqnarray}
\tau_{0} &=& \frac{5}{256\pi} f_{*}^{-1} \left(\frac{GM}{c^3}\pi  f_{*}\right)^{-5/3}\eta^{-1}\;,  \\
\tau_{1.5} &=& \frac{1}{8}f_{*}^{-1}\left(\frac{GM}{c^3}\pi  f_{*}\right)^{-2/3}\eta^{-1}\;, \\
M &=& m_1 + m_2 \;,\quad\mu = \frac{m_1m_2}{M}\;,\quad \eta =\frac{\mu}{M}\;,
\end{eqnarray}
where $f_*$, set to $10$~Hz in this paper, is the lower cutoff frequency of the detector arising from the steep rise in seismic noise.
(ii) The time $t_c$ at which the end of 
the inspiral signal arrives at the ECEF origin.
(iii) The phase of the signal at $t_c$ given by $\phi_c$.
 (iv) The overall amplitude of the signal denoted by $A$. The duration of the signal starting from the time at which its instantaneous frequency equals $f_\ast$ to $t_c$ is given by $\tau_0+\tau_1-\tau_{1.5}+\tau_2$ where,
\begin{eqnarray}
\tau_1 & = & \frac{5}{192\pi} f_{*}^{-1} \left(\frac{GM}{c^3}\pi f_{*}\right)^{-1}\eta^{-1}\left(\frac{743}{336} + \frac{11}{4}\eta\right)\;,\\
\tau_2 & = & \frac{5}{128\pi}f_{*}^{-1}\left(\frac{GM}{c^3}\pi  f_{*}\right)^{-1/3}\nonumber\\
&& \eta^{-1}\left(\frac{3058673}{1016064}+\frac{5429}{1008}\eta+\frac{617}{144}\eta^2 \right)\;,
\end{eqnarray}
are additional chirp times for the 2PN waveform.

\subsection{Likelihood ratio for a detector network}
Under our assumption of Gaussian, stationary noise, the log-likelihood Ratio (LLR)~\cite{Helstrom_95} for the ${i}^{\rm th}$ 
detector is given by, 
\begin{eqnarray}
    {\rm ln}\; \lambda^{{i}} &=& {\langle}x^{i}|h^{i}{\rangle} - \frac{1}{2}{\langle}h^{i}|h^{i}{\rangle}\;,
\end{eqnarray}
where, with  $\widetilde{a}(f)$ denoting the Fourier transform of any function $a(t)$ of time, 
\begin{eqnarray}
\label{eq.dot_product}
{\langle}\;p\; |\; q\;{\rangle}  &=& 4 \; {\rm Re} \int_{0}^{\infty} df\; 
\frac{\widetilde{p}(f)\widetilde{q}^\ast(f)} {{S_n}(f)}\;.
\end{eqnarray}
 
If we assume the noise in different detectors 
to be statistically independent, the log-likelihood for a $D$ detector network is given by,
\begin{eqnarray}
		\rm {ln}\;\lambda &=& \sum_{i=1}^{D}\left[ {\langle}x^i|h^i{\rangle} - \frac{1}{2}{\langle}h^i|h^i{\rangle}\right]\;.
\label{eq:NetworkLH}
\end{eqnarray}

It follows that, for a given data realization, the  log-likelihood is a function of the 
parameters $\theta = \{\tau_0,\tau_{1.5}, \alpha, \delta, t_c\}$, $A$, $\psi$, and $\phi_c$.

The GLRT statistic is  the global maximum of the LLR over
the parameters mentioned above. 
Adopting the notation in~\cite{bose_thilina}, we use the equivalent of the GLRT statistic 
defined as 
\begin{eqnarray}
\rho_{\rm coh}^2 & = & 2 \max_{A,\psi,\phi_c,\theta}\ln \lambda\;,
\end{eqnarray}
and call $\rho_{\rm coh}$ the {\it coherent search statistic}.

The
Maximization can be carried out as,
\begin{eqnarray}
\rho_{\rm coh}^2 & = &  \max_{\theta}\gamma^2(\theta)\;,\\
\gamma^2(\theta) & = & 2 \max_{A,\psi,\phi_c} \ln \lambda\;.
\end{eqnarray}
The inner maximization  can be performed analytically, while
the outer maximization  over  $\theta$ must be carried out numerically.

For fixed $\{\tau_0, \tau_{1.5}, \alpha, \delta\}$,
the maximization over  $t_c$ 
 can be carried out very efficiently using the 
Fast Fourier Transform (FFT)~\cite{bose_thilina}.  
The FFT-based approach is 
made especially convenient by the fact that the polarization waveforms can be 
generated directly in the Fourier domain using the stationary phase approximation~\cite{PhysRevD.44.3819}.
Thus, the outer maximization can be further split over $\Theta = \{\tau_0, \tau_{1.5}, \alpha, \delta\}$
and $t_c$.
We call 
\begin{eqnarray}
\Gamma^2(\Theta) & = & \max_{t_c} \gamma^2(\theta)\;\Rightarrow\;\rho^2_{\rm coh}=\max_\Theta\Gamma^2(\Theta),
\label{eq:Max_LLH_Analytical}
\end{eqnarray}
the {\it coherent fitness} function.
The computational cost of maximizing the coherent fitness over $\Theta$
 is the main challenge in the implementation of a fully
coherent search. 

In any practical implementation of the fully-coherent all-sky search,
the data streams from GW detectors must be analyzed in finite length
segments. Consequently, for fixed $\Theta$, $\gamma^2(\theta)$ is not valid for $t_c$  
greater than the length of the segment being analyzed. However, the FFT-based calculation 
generates $\gamma^2(\theta)$ for $t_c$ going up to the sum of the segment length and
the length of the signal corresponding to $\Theta$. To account for this effect, the spurious 
values of $\gamma^2(\theta)$ are deleted and consecutive 
data segments are overlapped by at least the length of the signal. In this paper, 
we analyze simulated data realizations that are generated independently of each other. 
Hence, there is no need to overlap them. However, care is taken to ensure that the 
signal embedded in the data has a $t_c$ that is sufficiently small. As long as the 
estimated $t_c$ does not stray too far from its true value, which was verified to be
the case in our simulations, ignoring deletion and overlapping does not affect our results.

\section{Particle Swarm Optimization}
\label{pso}

Although PSO started off as a single algorithm, the term now refers to a diverse set of algorithms and it  is viewed  more properly as a metaheuristic -- 
a general approach to optimization 
 based on a particular type of physical or biological model. In the case of 
PSO, the model happens to be
the flocking or swarming behavior of biological agents (birds, fish etc.). In fact, the PSO 
metaheuristic is one 
among many others that are inspired by nature. 

PSO  uses a fixed number of samples (called {\em particles}) of the
function to be optimized (called the {\em fitness function}). The 
particles move in the search space following stochastic iterative 
rules called the {\em dynamical equations}. 
In WM, the PSO algorithm used was
the global-best (or {\em gbest}) variant. In this paper,
we use the local-best (or {\em lbest}) variant of PSO described below
that leads to a significantly improved performance.

\subsection{The lbest PSO algorithm}
\label{pso_algorithm}
We adopt the following notation
in this section for describing lbest PSO.
\begin{itemize}
    \item $f(x)$: the scalar  fitness function to be maximized, with 
     $x = (x^1, x^2, \ldots, x^d)\in \mathbb{R}^d$. 
    In our case, $x = \Theta$, $f(x)$ is the coherent fitness 
    function $\Gamma^2(\Theta)$ (c.f., Eq.~\ref{eq:Max_LLH_Analytical})  and $d=4$.
    \item $\mathcal{S}\subset \mathbb{R}^d$: the search space
     defined by the hypercube $a^i\leq x^i \leq b^i$, $i = 1, 2, \ldots, d$.
     Among a set of locations in $\mathcal{S}$, the best location is the one with the maximum 
      fitness. 
    \item $N_p$: the number of particles in the swarm.
    \item $x_i[k]$: the position of the $i^{\rm th}$ particle
    at the $k^{\rm th}$ iteration.
    \item $p_i[k]$: the best location found by the $i^{\rm th}$ particle over all iterations up to and including the $k^{\rm th}$. $p_i[k]$ is called the {\em personal best} position
    of the $i^{\rm th}$ particle.
    \begin{eqnarray}
        f(p_i[k]) & = &\max_{j\leq k} f(x_i[j])\;.
    \end{eqnarray}
    \item $n_i[k]$: a set of particles, called the nearest neighbors of particle $i$, $n_i[k]\subseteq 
    \{1,2,\ldots,N_p\}\setminus \{i\}$. There are many possibilities, called {\em topologies}, for
       the choice of $n_i[k]$. In this paper, we use the ring topology with $2$
       neighbors in which
       \begin{eqnarray}
       n_i[k] = \left\{\begin{array}{cc}\{i-1,i+1\}\;,& i \notin \{1,N_p\}\\
                                        \{N_p,i+1\}\;, & i = 1\\
                                        \{i-1,1\}\;, & i = N_p
       \end{array}\right.\;.
       \end{eqnarray}
    \item $l_i[k]$: the best location among the particles in $n_i[k]$.
                    $l_i[k]$ is called the {\em local best} position of the $i^{\rm th}$ particle.
    \begin{eqnarray}
        f(l_i[k]) & = & \max_{j \in \{i\} \cup n_i[k]} f(x_j[k])\;.
    \end{eqnarray}
    \item $p_g[k]$: The best location among all the particles over all iterations up to and 
        including the $k^{\rm th}$.
        \begin{eqnarray}
        f(p_g[k]) & = & \max_{1\leq j \leq N_p} f(x_j)\;.
        \end{eqnarray}
\end{itemize}
The dynamical equations for {\em lbest} PSO are as follows.
\begin{eqnarray}
v_i[k+1] & = & w[k] v_i[k] + c_1 {\bf r}_1 (p_i[k] - x_i[k]) +\nonumber\\
            &&   c_2 {\bf r}_2 (l_i[k] - x_i[k])\;.
            \label{velocityEqn}\\
x_i[k+1] & = & x_i[k] + z_i[k+1]\;,\\
z^j_i[k] & = & \left\{
\begin{array}{cc}
  v_i^j[k] ,  &  -v_{\rm max}^j \leq v_i^j[k] \leq v_{\rm max}^j\\
   -v_{\rm max}^j ,  & 
   v_i^j[k] < -v_{\rm max}^j\\
   v_{\rm max}^j & v_i^j[k] > v_{\rm max}^j
\end{array}
\right.\;,
\end{eqnarray}
Here, $v_i[k]$ is called the ``velocity" of the $i^{\rm th}$ particle, $w[k]$ is a deterministic function known as the inertia weight (see below), $c_1$ and $c_2$
are constants, and ${\bf r}_i$ is a diagonal matrix with 
independent, identically distributed random variables having a uniform distribution 
over $[0,1]$. The second and third terms on the RHS of Eq.~\ref{velocityEqn} are 
called the {\it cognitive} and {\it social} terms respectively. 

The iterations are initialized at $k=1$ 
by independently drawing (i)  $x_i^j[1]$ 
from a uniform distribution over $[a^j,b^j]$, and (ii) 
$v_i^j[1]$ from a uniform distribution over $[-v_{\rm max}^j, v_{\rm max}^j]$.
The algorithm terminates when the number of 
iterations reaches a prescribed number ${N_{\rm iter}}$.
The solutions to the maximizer and maximum value of
the fitness found by PSO
are $p_g[N_{\rm iter}]$ and $f(p_g[N_{\rm iter}])$ respectively.

To handle particles that exit the
search space, we use the ``let them fly" 
boundary condition under which a particle outside the search space is 
assigned a fitness values of $-\infty$. Since both $p_i[k]$ and 
$l_i[k]$ are always within the search space, such a particle is
eventually dragged back into the search space by the 
cognitive and social terms.


The role of the
inertia weight, $w[k]$, is to control the degree of
exploration of the search space by allowing a particle 
to overcome the attractive cognitive and social terms.
In the version of PSO used here, the inertia weight $w[k]$ decreases linearly with $k$
from an initial value $w_{\rm max}$ to a final value $w_{\rm min}$. 
 Decreasing the inertia
weight transitions PSO from an initially exploratory to a final exploitative phase, and 
the longer the interval, namely $N_{\rm iter}$,
over which this happens, the later the onset of the 
transition. 
\subsection{PSO tuning metric}
\label{tuning}
As mentioned earlier, PSO is not guaranteed to 
converge to the global optimum of a fitness function and, as with any 
stochastic optimization algorithm, its parameters need to be tuned for it to perform well on a given fitness function. Fortunately, extensive 
studies have shown that most of the parameters of PSO can be set at 
near-fiducial values across a wide range of fitness functions~\cite{bratton+kennedy}. In WM, the
only parameter that was tuned was $N_{\rm iter}$, while the rest were fixed as follows: $N_p = 40$, 
$c_1 = c_2 =  2.0$,
$w_{\rm max} = 0.9$,
$w_{\rm min} = 0.4$, and $v_{\rm max} = 0.2$.

For a given $N_{\rm iter}$, the probability of 
convergence can be increased by the simple strategy of running 
multiple runs of PSO on the same data realization and choosing the 
best fitness value found across the runs. The probability of missing the global optimum decreases 
exponentially as $(1-P_{\rm conv})^{N_{\rm runs}}$, where $P_{\rm conv}$ is
the probability of successful convergence in any one run. 
This strategy was used 
in WM with an ad hoc choice for $\nruns$.
In this paper, we propose a metric-based objective process 
for tuning both $\nruns$ and  $\niter$.

The metric we use for tuning PSO follows from~\cite{pso_cbc} and is based on the fact that estimation error is caused by a shift of the 
 global maximum of the coherent fitness function
 away from the true signal parameters. Therefore, the minimum expectation 
 from any optimization method is that 
 the value found for the 
 coherent fitness  
  be higher than its value at the true signal parameters. 
  A failure of this condition indicates that  the global maximum
  of the coherent fitness function was not found.
  It is important to emphasize here that a failure in locating the global maximum 
does not necessarily mean a failure in detecting a signal. 
This point is discussed further in Sec.~\ref{sec:detection}.
  
 Stated formally, 
 let $\rho_{\rm coh}^\prime(\nruns, \niter)$  be the best  
  coherent fitness value found over $\nruns$ independent runs
of PSO, with $\niter$ iterations per PSO run. (We will occasionally 
drop $\nruns$ and $\niter$ and
simply use $\rho_{\rm coh}^\prime$ when there is no scope for confusion.)
Let $\rho^{(0)}_{\rm coh}$ be the coherent fitness value
for the true signal parameters. (Note that $\rho^{(0)}_{\rm coh}$ is a 
random variable due to noise in the data.)
The metric, denoted by $\mathcal{M}(\nruns,\niter)$,
is defined as    
\begin{eqnarray}
\mathcal{M}(\nruns,\niter) & = & 
{\rm Pr}\left(\rho_{\rm coh}^\prime(\nruns, \niter) \leq \rho^{(0)}_{\rm coh}\right)\;,
\label{eq:pso_tuning_metric}
\end{eqnarray}
The goal of tuning PSO should be to reduce $\mathcal{M}(\nruns,\niter)$ to an acceptable level, with $\mathcal{M}(\nruns,\niter) = 0$ being the most stringent requirement.

Performing multiple runs of PSO does not add to the execution time of the
overall search if the runs can be implemented in parallel. Hence, 
$N_{\rm runs}$ need not be tuned carefully if one has access to a sufficient
number of processors. In most situations, however, where the user of a supercomputer is billed by the number of hours and processors consumed, tuning $N_{\rm runs}$ can be beneficial. The tuning of $N_{\rm runs}$ is, of course, a definite requirement in situations where it affects the execution time of a search.

\section{Results}
\label{Results}
The performance of PSO is analyzed using simulated realizations of data, following the model
described in Sec~\ref{datamodel}, for the H, L, V, and K detector network.  

The simulated signals are normalized to have a specified {\it optimal network signal to noise ratio} (${\rm SNR_{{\rm opt}}}$), defined as,
\begin{eqnarray}
    {\rm SNR}_{{\rm opt}} 
    & = & {\left[\sum_{i=1}^D \langle h^i | h^i \rangle \right]^{1/2}}\;.
    \label{optimalSNR}
\end{eqnarray}
${\rm SNR}_{{\rm opt}}$ has the straightforward interpretation of being the ratio
of the mean of the LLR under $H_1$ to its standard deviation under $H_0$
in the case of binary hypotheses, where the waveform $h^{i}(t)$ is completely known under $H_1$. 
Although it does not numerically
match the corresponding ratio for the GLRT,
${\rm SNR}_{{\rm opt}}$ is still a convenient way to normalize signals because
the performance of GLRT depends  monotonically on it
and it admits a closed form expression. 

For the data realizations under
$H_1$,  we use the source locations L4 ($\alpha = 32.09^\circ$,
$\delta = -53.86^\circ$)
and L5 ($\alpha = 150.11^\circ$, 
$\delta = -60.16^\circ$)
from WM that correspond to the worst
and best 
condition numbers of the antenna pattern matrix. The signal waveforms used correspond to binaries with equal 
mass components, with the component mass labeled by M1 ($14.5$~$M_\odot$) and M2 ($1.506$~$M_\odot$). The signal lengths corresponding to M1 and M2 are $20.8$~sec and $23$~min respectively. 

Data realizations containing the
M1L{\em a}, $a = 4, 5$, signal, called M1 data, have a length of 64~sec, while those with the M2L{\em a} signal, called M2 data,
are 30~mins long. The sampling 
frequency of the data in all cases is 2048~Hz. The signal start times were fixed at 10~secs and 5~mins for M1 and M2 data respectively. 

The search space for PSO is fixed as follows. The range
for $\alpha$ and $\delta$ covers the entire sky.
For M1 data realizations, $\tau_0 \in [0, 50]$~sec and $\tau_{1.5}\in [0, 5]$~sec. For M2 data realizations, 
$\tau_0\in [500, 1000]$~sec and $\tau_{1.5}\in [5, 10]$~sec.

The results from the simulations are organized as follows.
First, we compute the PSO tuning metric $\mathcal{M}(\nruns,\niter)$ defined in 
Eq.~\ref{eq:pso_tuning_metric} for a discrete set of ${\rm SNR}_{{\rm opt}}$ values 
and obtain the optimum settings for $\nruns$ and $\niter$  
for each ${\rm SNR}_{{\rm opt}}$.
Next, the detection performance is quantified with these settings for each
${\rm SNR}_{{\rm opt}}$. From this, we obtain the minimum ${\rm SNR}_{{\rm opt}}$ at which there is almost complete separation of the distributions of 
$\rho_{\rm coh}^\prime(\nruns,\niter)$ under $H_0$ and $H_1$. The estimation performance of the 
PSO-based search is then quantified at this value of ${\rm SNR}_{{\rm opt}}$ and the associated optimum settings for $\nruns$ and $\niter$.
\subsection{PSO tuning}
\label{sec:tuning_results}
 
 The metric $\mathcal{M}(\nruns,\niter)$ is estimated using simulated data realizations.
 The computational burden of this estimation can be
 reduced substantially, following the strategy described below,
 by taking into account the fact that  
the  information needed for a given $\nruns$ is already nested within
that for a larger $\nruns$. 

For a given $\niter$ and ${\rm SNR}_{{\rm opt}}$, 
simulate $N_{\rm trials}$
 data realizations containing the same signal in each.
\begin{itemize}
    \item For each data realization, do $N_{\rm runs, max}$ independent PSO runs 
    and obtain the corresponding values of $\rho^\prime_{\rm coh}(1,\niter)$.
    \item For any $\nruns < N_{\rm runs, max}$, 
and for each data realization, draw
$N_{\rm btstrp}$ bootstrap~\cite{bootstrap} samples of size $\nruns$ from the
 set of $N_{\rm runs, max}$ values. 
    \item For each bootstrap sample, obtain the best fitness value 
among $\nruns$ runs, yielding $N_{\rm btstrp}$ values of $\rho^\prime_{\rm coh}(\nruns,\niter)$. 
\item Estimate
$\mathcal{M}(\nruns,\niter)$ for each realization as the fraction of bootstrap samples
for which $\rho^\prime_{\rm coh}(\nruns,\niter)\leq \rho^{(0)}_{\rm coh}$. 
\end{itemize}
Finally, from the $N_{\rm trials}$ value of $\mathcal{M}(\nruns,\niter)$ thus 
obtained, estimate suitable
statistical summaries. By repeating the above process for different 
values of $\niter$ and $\nruns < N_{\rm runs, max}$, one can pick the best
combination based on
some preset thresholds on the statistical summaries and available 
computational resources for carrying out the PSO-based search.

In this paper, we tune the PSO-based search for a given ${\rm SNR}_{{\rm opt}}$ using $N_{\rm trials} = 120$, $N_{\rm runs, max} = 12$, and $N_{\rm btstrp}=1000$. 
Table~\ref{Table:tuning}
shows statistical summaries -- the 1st, 50th (median), and 99th percentiles --
of the distribution of 
$\mathcal{M}(\nruns,\niter)$ for a discrete set of 
${\rm SNR}_{{\rm opt}}$ values. 
\begin{table*}
\caption{
The PSO tuning metric $\mathcal{M}(\nruns,\niter)$ 
for a discrete set of 
${\rm SNR}_{{\rm opt}}$ values. For each $\nruns$ and $\niter$ combination, there are  
three rows corresponding (from top to bottom) to ${\rm SNR}_{{\rm opt}} = 9$, $10$, and $11$ respectively. In each row, the numbers from left to right are the 1st percentile, Median, and 99th percentile of the distribution of 
$\mathcal{M}(\nruns,\niter)$.  
\label{Table:tuning}}
\centering
\begin{tabular}{| c | c | c | c | c | }
\hline
$\niter$ & $\nruns=1$ & $\nruns=2$ & $\nruns=3$ & $\nruns=4$ \\
\hline
\hline
500 & $\begin{array}{lcr}0.050 & 0.100 & 0.150\\
\hline
0.025 & 0.058 & 0.108\\
\hline
0.008 & 0.033 & 0.083\\
\end{array}$
 & $\begin{array}{lcr}0 & 0.025 & 0.067\\
\hline
0 & 0.008 & 0.033\\
\hline
0 & 0 & 0.017\\
\end{array}$
 & $\begin{array}{lcr}0 & 0.008 & 0.033\\
\hline
0 & 0 & 0.017\\
\hline
0 & 0 & 0.008\\
\end{array}$
 & $\begin{array}{lcr}0 & 0 & 0.017\\
\hline
0 & 0 & 0.008\\
\hline
0 & 0 & 0.008\\
\end{array}$
\\
\hline
1000 & $\begin{array}{lcr}0.008 & 0.042 & 0.075\\
\hline
0 & 0.017 & 0.058\\
\hline
0 & 0.008 & 0.033\\
\end{array}$
 & $\begin{array}{lcr}0 & 0.008 & 0.025\\
\hline
0 & 0 & 0.008\\
\hline
0 & 0 & 0.008\\
\end{array}$
 & $\begin{array}{lcr}0 & 0 & 0.008\\
\hline
0 & 0 & 0\\
\hline
0 & 0 & 0\\
\end{array}$
 & $\begin{array}{lcr}0 & 0 & 0.008\\
\hline
0 & 0 & 0\\
\hline
0 & 0 & 0\\
\end{array}$
\\
\hline
1500 & $\begin{array}{lcr}0 & 0.025 & 0.067\\
\hline
0 & 0.017 & 0.042\\
\hline
0 & 0 & 0.017\\
\end{array}$
 & $\begin{array}{lcr}0 & 0 & 0.017\\
\hline
0 & 0 & 0.008\\
\hline
0 & 0 & 0.008\\
\end{array}$
 & $\begin{array}{lcr}0 & 0 & 0.008\\
\hline
0 & 0 & 0\\
\hline
0 & 0 & 0\\
\end{array}$
 & $\begin{array}{lcr}0 & 0 & 0\\
\hline
0 & 0 & 0\\
\hline
0 & 0 & 0\\
\end{array}$
\\
\hline
2000 & $\begin{array}{lcr}0 & 0.025 & 0.067\\
\hline
0 & 0.008 & 0.025\\
\hline
0 & 0 & 0.025\\
\end{array}$
 & $\begin{array}{lcr}0 & 0 & 0.017\\
\hline
0 & 0 & 0.008\\
\hline
0 & 0 & 0\\
\end{array}$
 & $\begin{array}{lcr}0 & 0 & 0.008\\
\hline
0 & 0 & 0\\
\hline
0 & 0 & 0\\
\end{array}$
 & $\begin{array}{lcr}0 & 0 & 0\\
\hline
0 & 0 & 0\\
\hline
0 & 0 & 0\\
\end{array}$
\\
\hline
\end{tabular}
\end{table*}

For each ${\rm SNR}_{{\rm opt}}$, we pick the first
pair of $\nruns$ and $\niter$ at which all the three percentiles become zero,
implying that the estimated $\mathcal{M}(\nruns,\niter)=0$ for all $N_{\rm trials}$
realizations. 
Where the choice is between $\nruns=3$ and $\nruns=4$, we pick the latter 
as it is the more conservative option. As noted earlier, picking a 
slightly larger $\nruns$ does not necessarily increase the execution time of the search since the independent PSO runs
can be conducted in parallel (given the resources). Thus, the combinations $(\nruns,\niter)$ used
 are $(4,1500)$ and $(4,1000)$ for ${\rm SNR}_{{\rm opt}}=9$  and ${\rm SNR}_{{\rm opt}} \in \{ 10, 11\}$ respectively.

 The metric $\mathcal{M}(\nruns,\niter)$ used in the tuning 
 is estimated from a finite number of data realizations. As such, there is a non-zero 
 probability of $\mathcal{M}\neq 0$ even if it is 
 estimated to be zero in the tuning. To validate the tuning process, therefore, we test the performance of the tuned PSO over
 a much larger 
number of data realizations.

\begin{figure}
\centering
\includegraphics[width=0.5\textwidth]{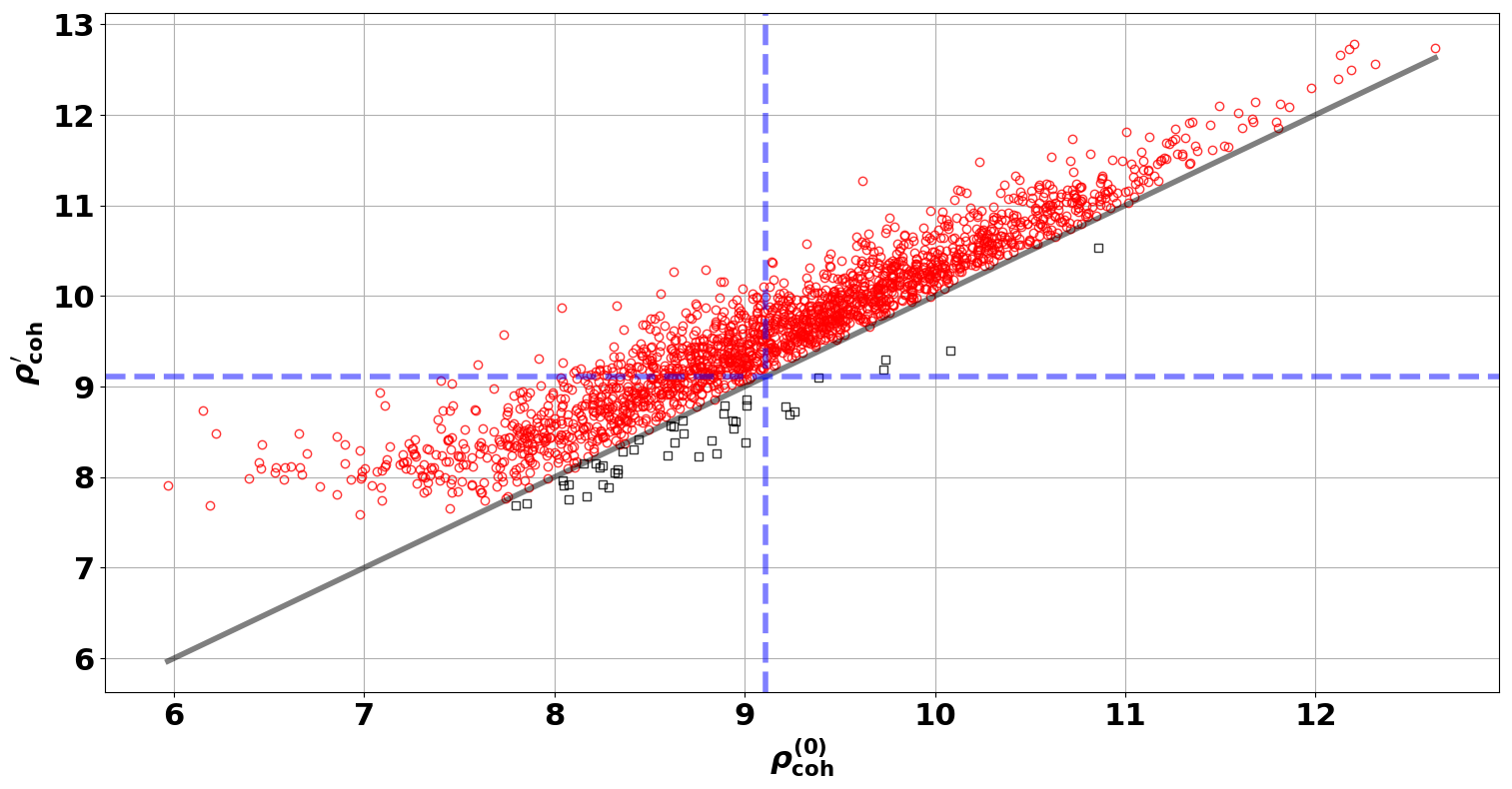}
\caption{\label{fig:aligo_css_vs_true_css_9}
Scatterplot of the best coherent fitness value, $\rho_{\rm coh}^\prime(4,1500)$, found by PSO and the coherent fitness at the true signal parameters, $\rho_{\rm coh}^{(0)}$, for ${\rm SNR}_{\rm opt}=9$ and M1 data. 
Out of a total of 2000 points, 44 (or $2.2\%$) fall below the diagonal. The blue lines show the detection threshold of 9.109.
}
\end{figure}

\begin{figure}
\centering
\includegraphics[width=0.5\textwidth]{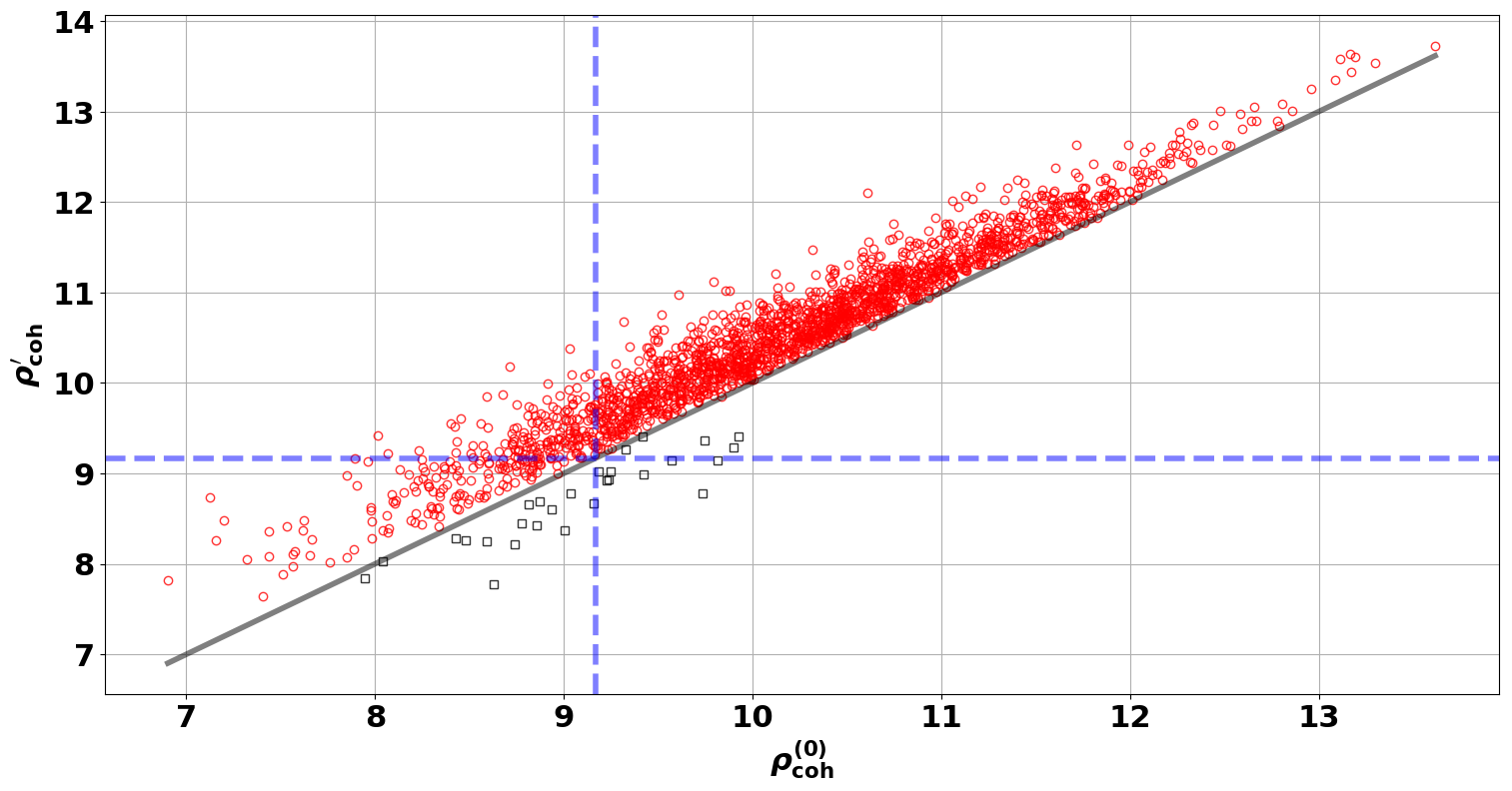}
\caption{\label{fig:aligo_css_vs_true_css_10}
Scatterplot of the best coherent fitness value, $\rho_{\rm coh}^\prime(4,1000)$, found by PSO and the coherent fitness at the true signal parameters, $\rho_{\rm coh}^{(0)}$, for ${\rm SNR}_{\rm opt}=10$ and M1 data. 
Out of a total of 2000 points, 28 (or $1.4\%$) fall below the diagonal. The blue lines show the detection threshold of 9.168.
}
\end{figure}

\begin{figure}
\centering
\includegraphics[width=0.5\textwidth]{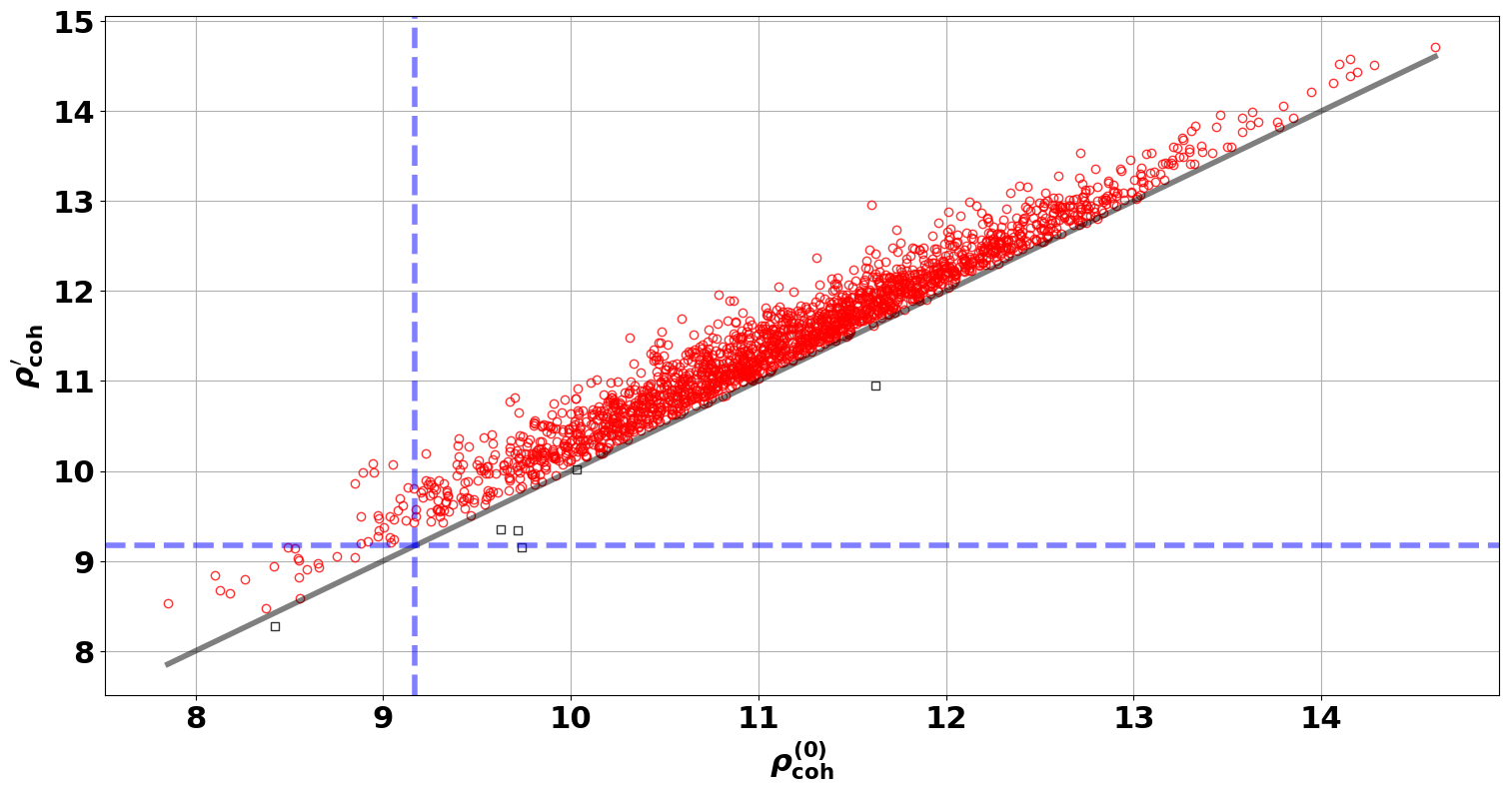}
\caption{\label{fig:aligo_css_vs_true_css_11}
Scatterplot of the best coherent fitness value, $\rho_{\rm coh}^\prime(4,1000)$, found by PSO and the coherent fitness at the true signal parameters, $\rho_{\rm coh}^{(0)}$, for ${\rm SNR}_{\rm opt}=11$ and M1 data. 
Out of a total of 2000 points, 6 (or $0.3\%$) fall below the diagonal. The blue lines show the detection threshold of 9.168.
}
\end{figure}

Figures~\ref{fig:aligo_css_vs_true_css_9} to \ref{fig:aligo_css_vs_true_css_11} show the 
validation of tuning for ${\rm SNR}_{\rm opt} = 9$, $10$, and $11$ respectively.
For each ${\rm SNR}_{\rm opt}$, there are $1000$ realizations of M1 data  for each 
of the two sky locations (L4 and L5).
The figures show the scatterplot of the best coherent fitness value, $\rho_{\rm coh}^\prime(\nruns,\niter)$, found by PSO and the coherent fitness at the true signal parameters, $\rho_{\rm coh}^{(0)}$, where $\nruns$ and $\niter$ are set to their tuned values.

As expected, the condition $\rho_{\rm coh}^\prime \geq \rho_{\rm coh}^{(0)}$ fails in a
non-zero fraction of the data realizations. However, 
the observed performance is a 
substantial improvement 
over gbest PSO in WM, where this condition 
failed in $6.6\%$ of the realizations at  ${\rm SNR}_{\rm opt}=12.7$
and $\nruns = 12$. Here, the same condition fails in $2.2\%$, $1.4\%$, and $0.3\%$ of trials for ${\rm SNR}_{\rm opt} = 9$, $10$,
and $11$ respectively.
Interestingly, the total number of fitness
evaluations ($N_p\times N_{\rm iter}\times N_{\rm runs}$) per data realization, which 
was found to be $2.4\times 10^5$ for gbest PSO in WM, remains the same for lbest
PSO despite the lower ${\rm SNR}_{\rm opt}$ of $9$. The number of fitness evaluations
shrinks, compared to gbest PSO, to $1.6\times 10^5$ at the higher ${\rm SNR}_{\rm opt}$ values. 

Interestingly, out of the $44$, $28$, and $6$ points that fall below the diagonal in 
Fig.~\ref{fig:aligo_css_vs_true_css_9} to Fig.~\ref{fig:aligo_css_vs_true_css_11} respectively,
$42$, $28$, and all $6$ arise from data realizations associated with the L5 location. 
 We do not understand the origin of this effect at present. The L5 location also seems to be 
 associated with  a 
prominent secondary maximum, discussed in Sec.~\ref{sec:estimation}, 
of the coherent fitness function. It is likely that these effects are linked.

The tuning process described above pertains to the specific length of M1 data. However, the optimum settings for a given data length 
can serve as useful starting points for other cases. We illustrate
this with the challenging case of M2 where the data length is $\approx 30$ times longer than
that for M1. Since the current computation time for the M2 data is extremely long ($\approx 18$ hours), carrying out a statistically reliable tuning with a sufficiently large number of data realizations is not feasible. Instead, we simply start with the optimum settings found for a given ${\rm SNR}_{\rm opt}$ in 
Table~\ref{Table:tuning} and explore variations around those settings. Considering the case of ${\rm SNR}_{\rm opt} = 11$ and starting with the tuned settings
($\nruns = 4$ and $\niter = 1000$) for M1 data, we  
found that the performance of PSO was not acceptable. However, by
increasing $\niter$ to 1500 and $\nruns$ to $8$ a reasonably effective performance level was achieved.

\begin{figure}
\centering
\includegraphics[width=0.5\textwidth]{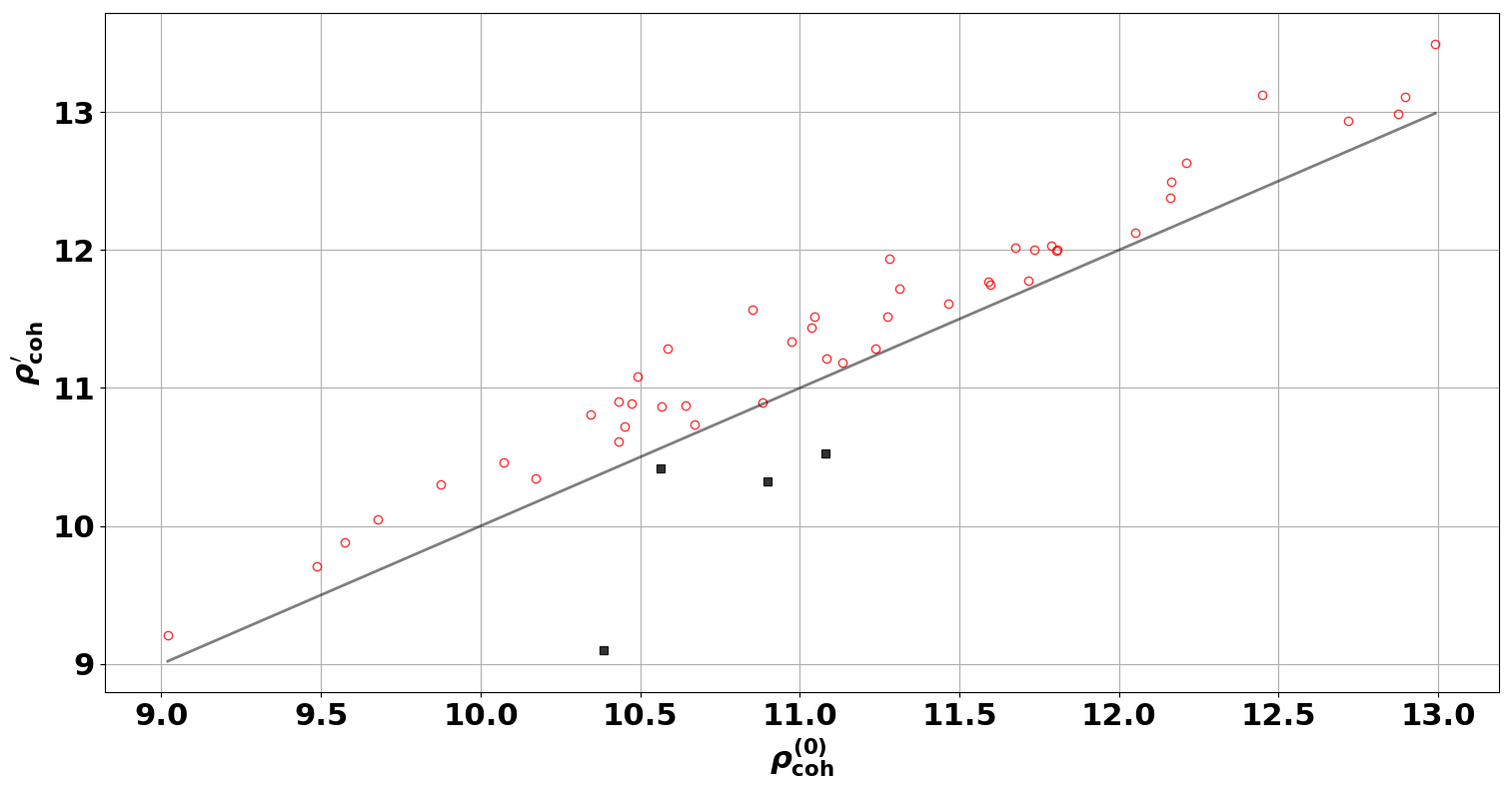}
\caption{\label{fig:aligo_css_scatter_snr11_LONG}
Scatterplot of the best coherent fitness value, $\rho_{\rm coh}^\prime(8,1500)$, found by PSO and the coherent fitness at the true signal parameters, $\rho_{\rm coh}^{(0)}$, for ${\rm SNR}_{\rm opt}=11$ and
M2 data. 
Out of a total of 50 points, 4 (or $8\%$) fall below the diagonal. 
}
\end{figure}

Fig.~\ref{fig:aligo_css_scatter_snr11_LONG} shows the scatterplot
of $\rho^\prime_{\rm coh}$ and $\rho_{\rm coh}^{(0)}$
for  data realizations containing the M2 signal with ${\rm SNR}_{\rm opt}=11$.
There are $25$ realizations of M2 data  for each 
of the two sky locations (L4 and L5).
The fraction of
trials in which the condition $\rho^\prime_{\rm coh} \geq \rho_{\rm coh}^{(0)}$ fails is now at $8\%$. 
As observed in the case of M1 data, these dropouts again arise from the data
realizations associated with the L5 location. The fraction of failures should go down 
with an increase in ${\rm SNR}_{\rm opt}$ and further improvements 
in performance should be possible with some more exploration of $\nruns$ and $\niter$.

\subsection{Detection performance}
\label{sec:detection}
Having obtained the best settings for PSO following the tuning strategy described in Sec.~\ref{sec:tuning_results},
we can characterize the detection and estimation 
performance that can be expected from the PSO-based search.
This is done almost exclusively for the M1 signal in this paper
because 
our current computing resource limitations do not allow
a reasonably large number of data realizations, 
especially under $H_0$, to be used for the M2 signal. 
For M1, we use the same data realizations that were used
in the validation (c.f., Fig.~\ref{fig:aligo_css_vs_true_css_9} to
Fig.~\ref{fig:aligo_css_vs_true_css_11}) of tuned settings.  
 The
number of noise-only data realizations under $H_0$ is $1000$ 
for each of the two $\nruns$ and $\niter$ combinations
obtained from the
tuning process.

\begin{figure}
\centering
\includegraphics[width=0.5\textwidth]{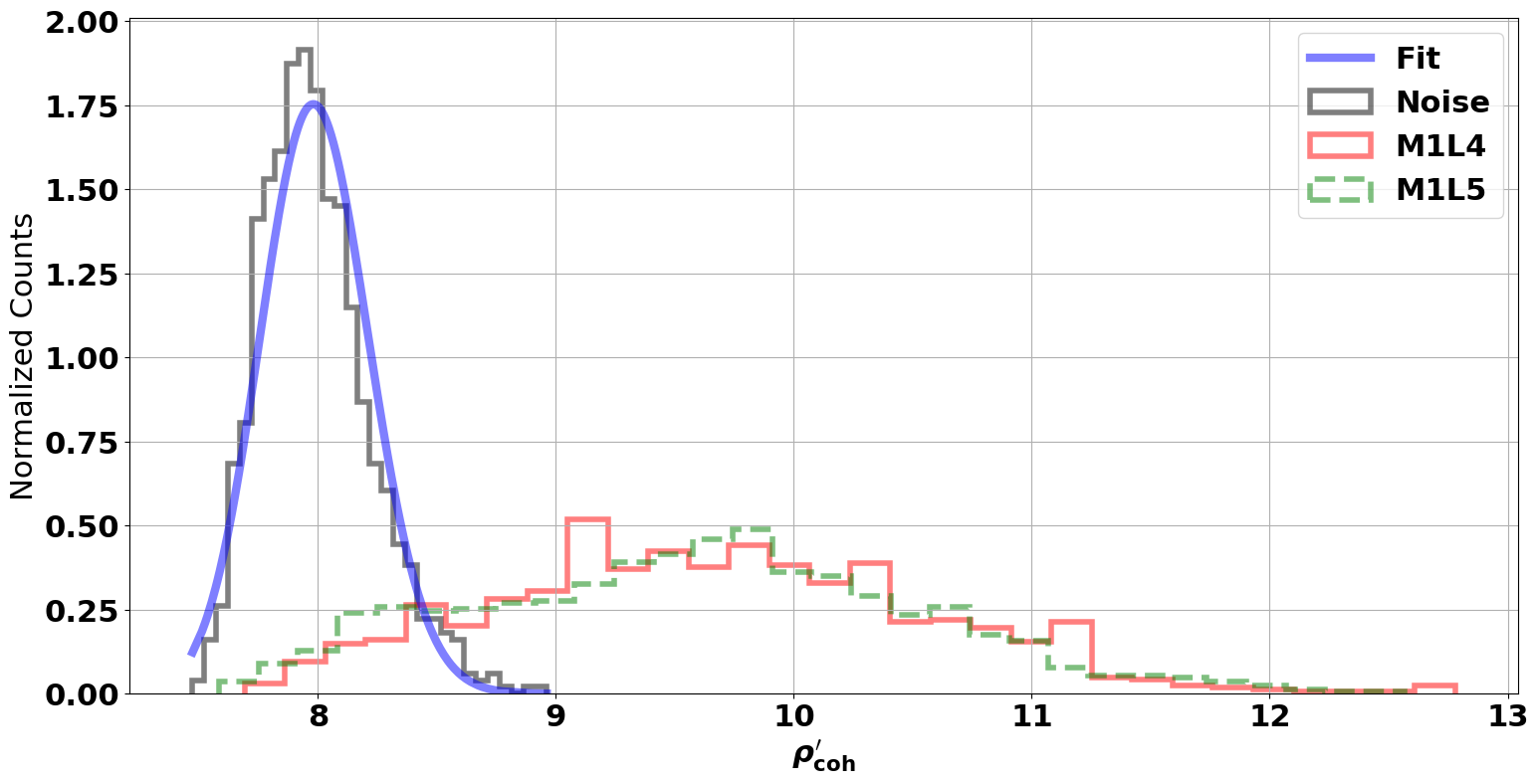}
\caption{\label{fig:aligo_css_hist_snr9}Histograms of the coherent search statistic found by PSO, $\rho^\prime_{\rm coh}(4,1500)$, for $H_0$ and $H_1$ data realizations. The latter contain the M1 signal
with
${\rm SNR}_{\rm opt}=9$  at two different sky locations (L4 and L5).  
The p-value of a two-sample 
Kolmogorov-Smirnov test between the samples corresponding to L4 and L5 is $0.13$.
}
\end{figure}

\begin{figure}
\centering
\includegraphics[width=0.5\textwidth]{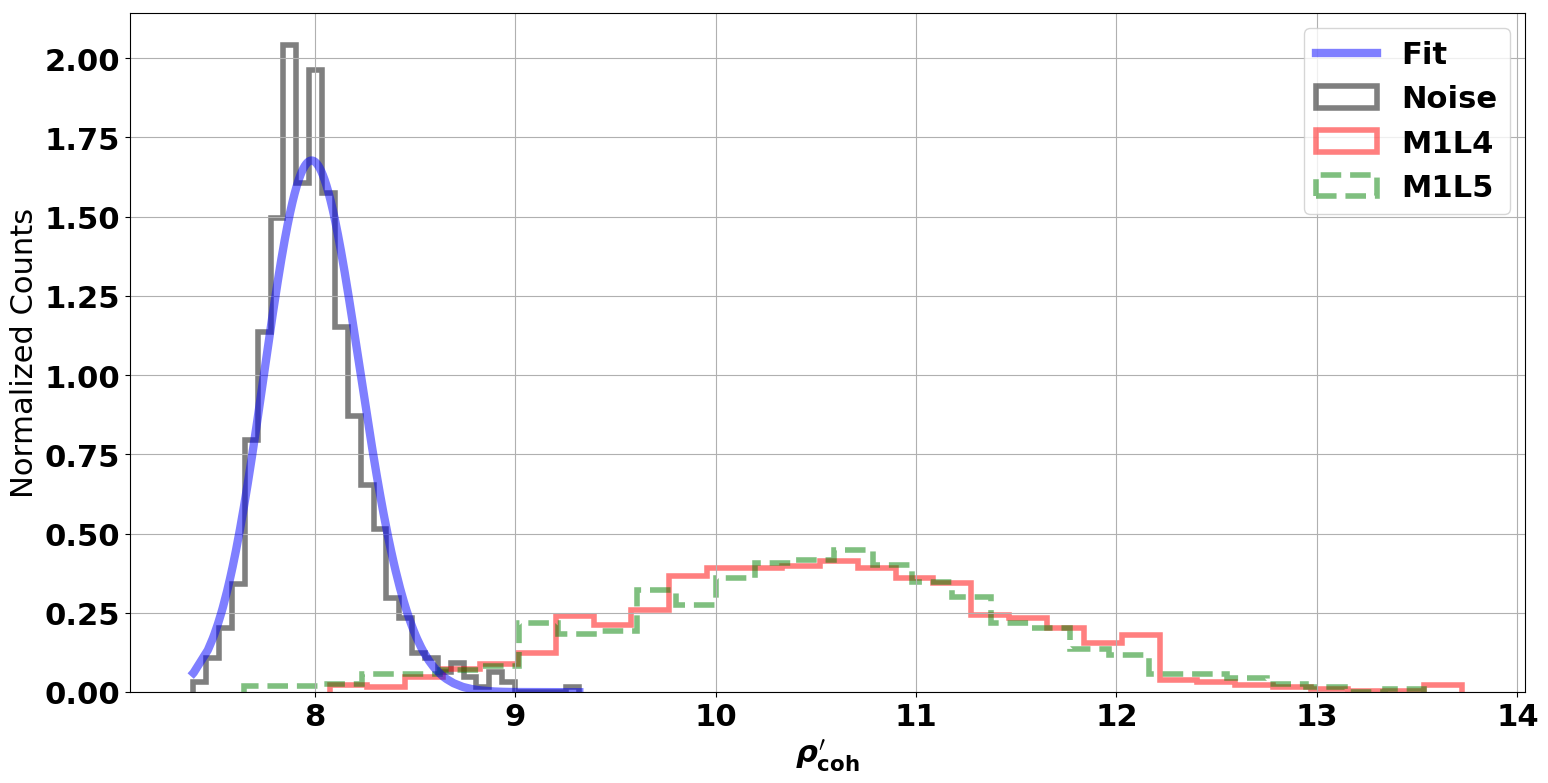}
\caption{\label{fig:aligo_css_hist_snr10}
Histograms of the coherent search statistic found by PSO, $\rho^\prime_{\rm coh}(4,1000)$, for $H_0$ and $H_1$ data realizations. The latter contain the M1 signal
with
${\rm SNR}_{\rm opt}=10$  at two different sky locations (L4 and L5).  
The p-value of a two-sample 
Kolmogorov-Smirnov test between the samples corresponding to L4 and L5 is $0.46$.
}
\end{figure}

\begin{figure}
\centering
\includegraphics[width=0.5\textwidth]{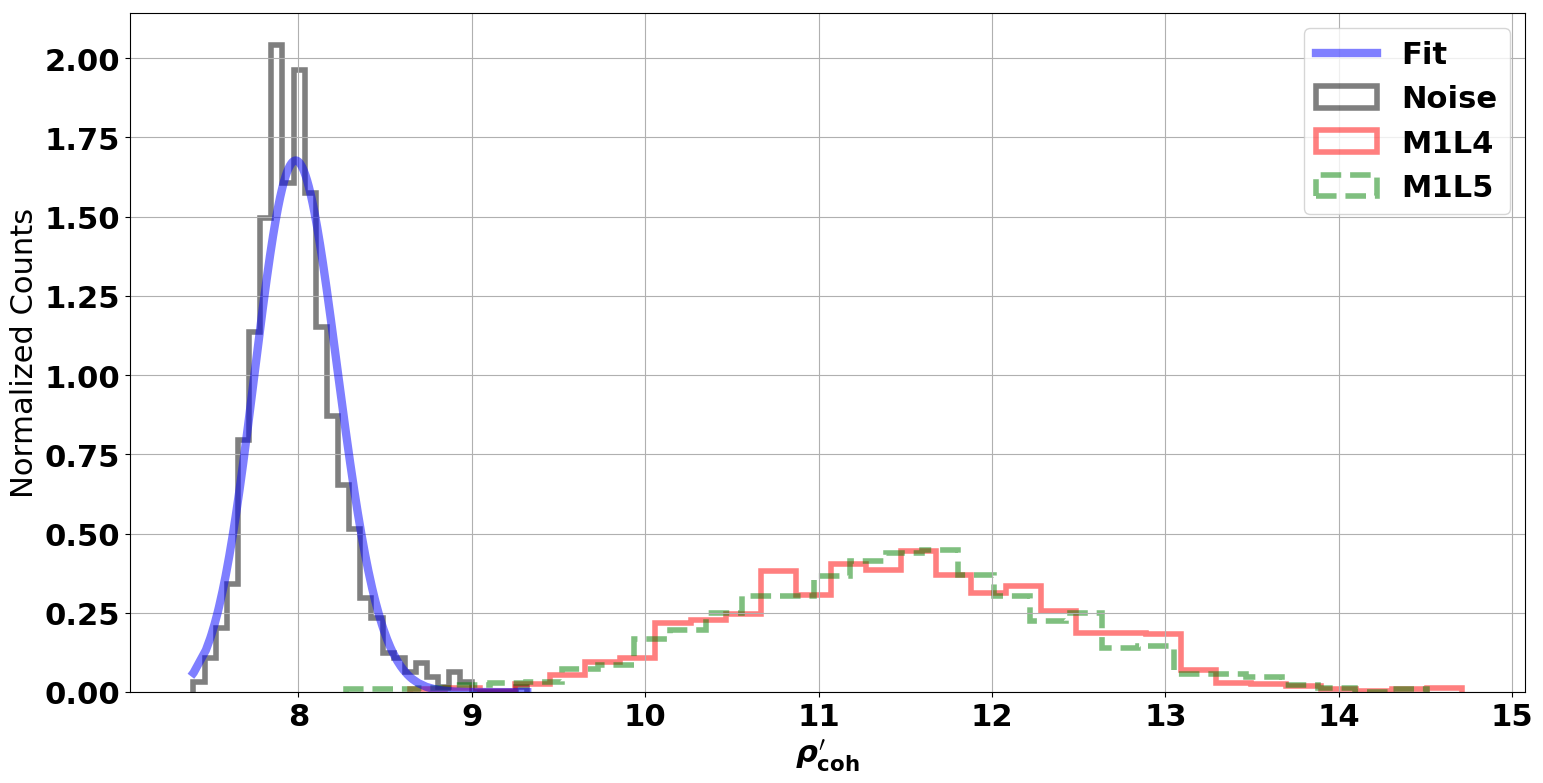}
\caption{\label{fig:aligo_css_hist_snr11}
Histograms of the coherent search statistic found by PSO, $\rho^\prime_{\rm coh}(4,1000)$, for $H_0$ and $H_1$ data realizations. The latter contain the M1 signal
with
${\rm SNR}_{\rm opt}=11$  at two different sky locations (L4 and L5). 
The p-value of a two-sample 
Kolmogorov-Smirnov test between the samples corresponding to L4 and L5 is $0.75$.
}
\end{figure}

Figures~\ref{fig:aligo_css_hist_snr9},~\ref{fig:aligo_css_hist_snr10}, and~\ref{fig:aligo_css_hist_snr11} show the distributions of the 
coherent search statistic found by PSO for the different data sets described above.  A two-sample Kolmogorov-Smirnoff (KS) test is 
performed on the samples associated with the two  sky locations.
In all cases, the test supports the null hypothesis that the 
two samples are drawn from the same distribution. This shows that,
as happens in the case of a grid-based search,
the distribution of the 
coherent search statistic found by PSO depends only on ${\rm SNR}_{\rm opt}$ and not the details of the individual detector 
responses that vary with the source location. 









    

We fit a lognormal probability density function (pdf) to the distribution of the 
coherent search statistic under $H_0$ and obtain the 
detection threshold from this fit for a given false alarm
probability (FAP). For a ${\rm FAP}= 2.03\times10^{-6}$,
which corresponds to a false alarm rate (FAR) of $1$~event/year, 
the detection thresholds for the two different PSO settings, $(\nruns = 4,\niter = 1500)$ and $(\nruns = 4, \niter = 1000)$, 
differ only marginally at $9.109$ and $9.168$ respectively.
Table \ref{Table:detection} lists the 
detection probabilities at these thresholds for different ${\rm SNR}_{\rm opt}$ values.
We see that the sensitivity of 
a PSO-based fully-coherent all-sky search reaches an interesting level
at around ${\rm SNR}_{\rm opt} \approx 9$. 



\begin{table}
\caption{
Detection probabilities for the M1 signal at a FAR of 1 false event per year (${\rm FAP}=2.03\times10^{-6})$. 
Also listed is the loss in detection probability, $L_{\rm DP}$,
defined in Eq.~\ref{eq:loss_detprob}.
\label{Table:detection}}
\centering
\begin{tabular}{| c | c | c | c | }
\hline
 ${\rm SNR}_{\rm opt}$ & L4 & L5 & $L_{\rm DP}$ \\
\hline
\hline
$9$ & 0.72 & 0.692 & $0.39\%$\\
\hline
$10$ & 0.934 & 0.903 & $0.48\%$ \\
\hline
$11$ & 0.995 & 0.985 & $0.05\%$\\
\hline
\end{tabular}
\end{table}

As discussed in Sec.~\ref{sec:tuning_results},
PSO has a finite probability of not 
converging to the global maximum of the coherent fitness function.
This is shown by the points that fall below the diagonal in 
Fig.~\ref{fig:aligo_css_vs_true_css_9} to Fig.~\ref{fig:aligo_css_scatter_snr11_LONG}. However, failure to converge to the global maximum does not necessarily mean failure to
detect a signal since the coherent search statistic, $\rho_{\rm coh}^\prime$, obtained from PSO  may still exceed the detection threshold. Detection probability is only reduced when 
$\rho_{\rm coh}^\prime$ falls both below the diagonal and below
the detection threshold. 

We define the loss in detection 
probability as
\begin{eqnarray}
L_{\rm DP} & = & P(\rho^\prime_{\rm coh}\leq \eta | \rho^{(0)}_{\rm coh}\geq \eta )\;,
\label{eq:loss_detprob}
\end{eqnarray}
where $P(A|B)$ denotes the conditional probability of event $A$
given event $B$, and $\eta$ is the detection threshold.
As seen from Table~\ref{Table:detection},
the estimated $L_{\rm DP}$ is negligible in all cases
for the M1 signal at a FAR of 1~false event per year. 

As mentioned earlier, we do not have a sufficiently large number of trials under $H_0$ for M2 data
to reliably estimate the detection threshold for a realistically small
FAP. Instead, we simply pick an ad hoc range of $9.0\leq \eta \leq 11.0$ for the detection threshold in this case and find that   $L_{\rm DP}$ 
varies between $2\%$ and  $6\%$.




\subsection{Estimation performance}
\label{sec:estimation}
Most analyses~\cite{abbott2018prospects,veitch2015parameter} 
of parameter estimation for the fully-coherent all-sky CBC search report 
 Bayesian credible regions. A credible region ${\rm CR}_{\alpha}$ is the smallest volume 
 in signal
 parameter space that encloses
 a fraction $\alpha$ of the total posterior probability.  As such, ${\rm CR}_{\alpha}$
 is derived from a single data realization. It should be emphasized that  
 ${\rm CR}_{\alpha}$ is in general not equivalent to a Frequentist confidence region in that it 
 is not guaranteed to cover the true parameter values with a probability of $\alpha$ over
 multiple data realizations. 
 The reduced computational cost of the PSO-based search makes it 
 feasible to obtain Frequentist errors for point parameter estimates
 using multiple data realizations.
 
In this paper, we do a limited examination of parameter estimation errors by
focusing on only 
the M1 signal at ${\rm SNR}_{\rm opt}=11$ for which its
detection probability is nearly unity. A more thorough examination of this topic 
will be the subject of future work.

\begin{figure}[htb!]
\centering
\includegraphics[width=0.5\textwidth]{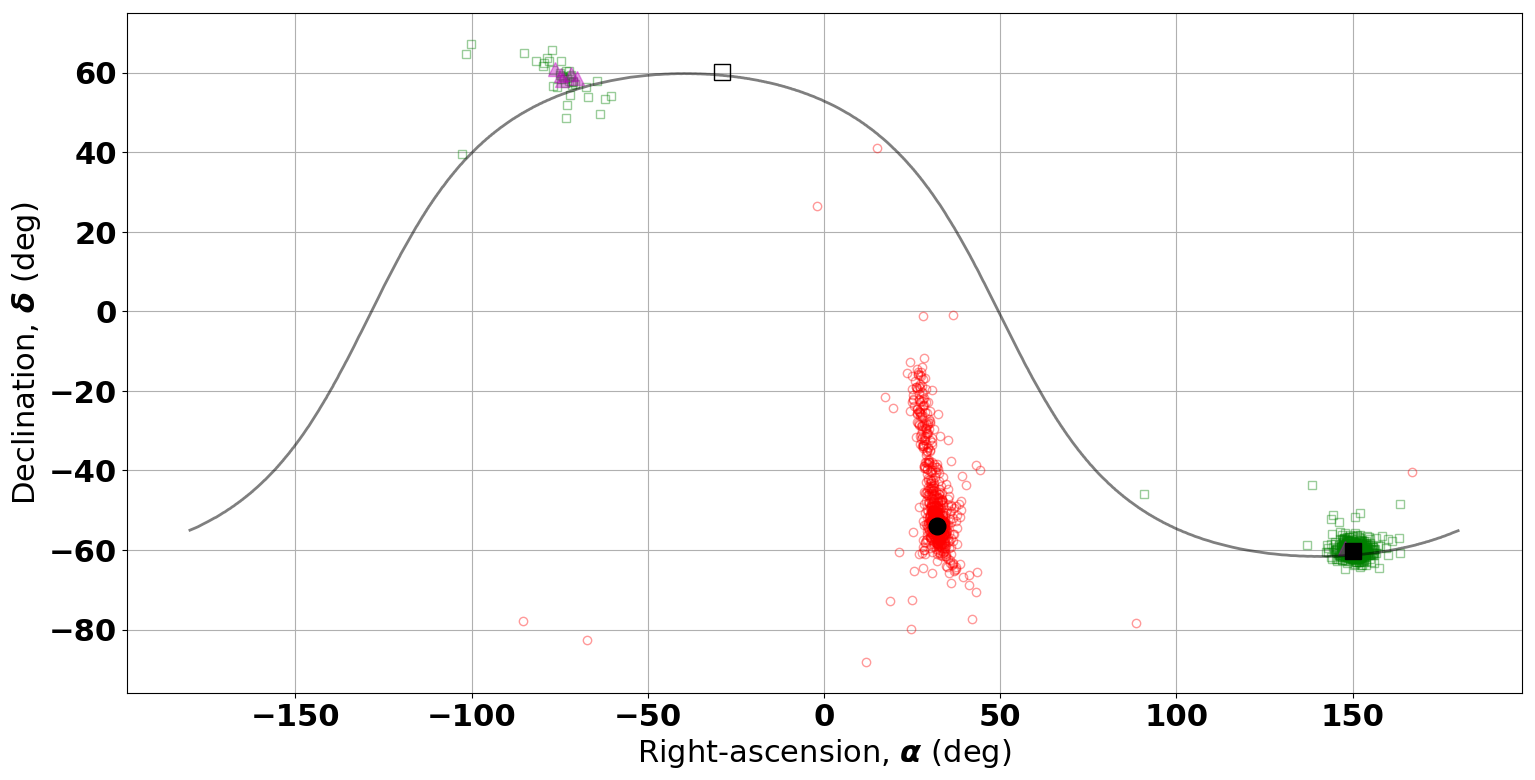}
\caption{\label{fig:aligo_sky_map}Estimated sky locations for 
a ${\rm SNR}_{\rm opt}=11$ M1 signal at the L4 (red) and L5 (green and magenta) sky locations. The true sky locations for L4 and L5 are denoted by the filled 
black circle and square respectively. All $6$ of the data
realizations for which the condition $\rho_{\rm coh}^\prime \ge \rho_{\rm coh}^{(0)}$ failed, indicating failure in finding the global maximum of the coherent fitness function, are associated with the L5 location and are marked by magenta triangles. The secondary maximum associated with L5 lies along the 
great circle (solid curve) joining L5 with its antipodal point (empty black square).
}
\end{figure}

Fig.~\ref{fig:aligo_sky_map} shows the estimated locations
on the sky for both L4 and L5. As was observed in WM for the case of
initial LIGO PSD, the error in sky location generally follows
its dependence on the condition number of the antenna pattern 
matrix: the estimated locations are distributed over a much more
extended region when the source location has a higher condition number. 

A caveat to the empirical rule of thumb above 
is that the L5 location, while 
exhibiting a more compact distribution of the estimates, also 
shows a widely separated secondary maximum. Going 
by the condition $\rho^\prime_{\rm coh} \geq \rho^{(0)}_{\rm coh}$  as
an indicator of convergence to the global maximum, we find that
 the global maximum shifted to the secondary in $4.3\%$ of the trials
 where this condition was met. At the same time,  all but one of the trials where
this condition failed also
appear at the secondary maximum. Thus, PSO
 does not go to an arbitrary location but latches on to the secondary when it fails to locate
 the global maximum. These observations indicate that 
the secondary maximum is a genuine feature of the coherent fitness function when a source
is located at L5 and not an artifact of using PSO.

\begin{figure}[htb!]
\centering
\includegraphics[width=0.5\textwidth]{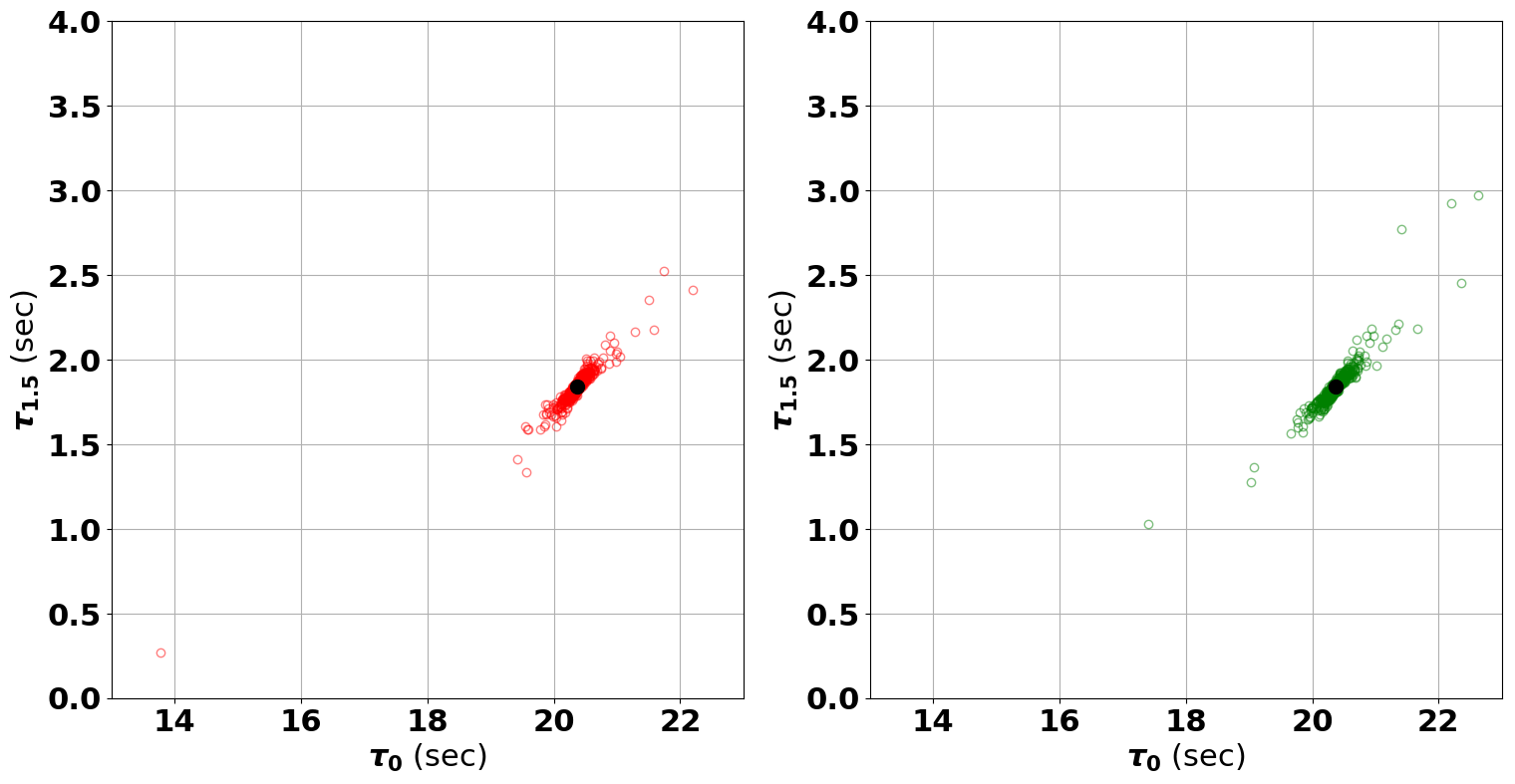}
\caption{\label{fig:aligo_chirp_map}Shown are the estimated chirp times for 1000 realizations of a $\rho_{\rm coh}^{(0)}=11$ signal at the M1L4 and M1L5 sky locations using $\rho_{\rm coh}^\prime(4,1000)$. The true chirp time is marked by a black circle for L4 and L5, respectively.}
\end{figure}

Fig.~\ref{fig:aligo_chirp_map} shows the estimation error 
distribution for chirp times $\tau_0$ and $\tau_{1.5}$. The main
observation here is that these errors are essentially uncorrelated
with the errors in sky localization. This is expected from a 
Fisher information analysis but, as was pointed out in WM, 
this need not be true for some sky locations.

\section{Computation time analysis}
\label{sec:latency}
As demonstrated by the rich science payoffs from the joint GW and EM 
observations of GW170817, the detection of EM counterparts of 
CBC events is of critical importance. A key factor 
here is the time 
elapsed in the 
dissemination of a follow up alert to EM astronomers. 

While the time needed to generate an alert depends on a number of factors, including 
manual vetting of detector state and data quality,  the execution 
time of the GW search algorithm is the primary
determinant. With this in mind,
it is important to analyze the computation time of a PSO-based fully coherent search.

The results in this paper are obtained
using a code that is implemented in the C programming language. 
There are two nested layers of parallelization in the code.
In the outer layer, parallelization is performed over
the independent PSO runs, while the inner layer 
parallelizes over the  PSO particles. 

The outer parallelization does not 
involve any communication between the parallel processes. It is implemented 
using a framework for high throughput computing 
called \textsc{launcher}~\cite{a2017launcher} that allows  
running large collections of independent applications
on batch-scheduled clusters. 
The inner parallelization is implemented 
as a multi-threaded application based on OpenMP~\cite{Dagum:1998:OIA:615255.615542,openmp08}, which assigns the fitness calculation for each particle to a unique thread that acts as an effectively independent processing
unit. 

On the cluster used for obtaining the results in this paper (TACC Lonestar 5 \cite{tacc_ls5}), each independent PSO run is assigned 
to a separate node, with each node 
providing up to 48 independent threads (24 processing 
cores with 2 hardware threads per core). On a given node we assign one thread to each particle, which allows for a parallel calculation of the   
$N_p=40$ particle fitness values computed in each iteration of PSO.

\begin{figure}
\centering
\includegraphics[width=0.5\textwidth]{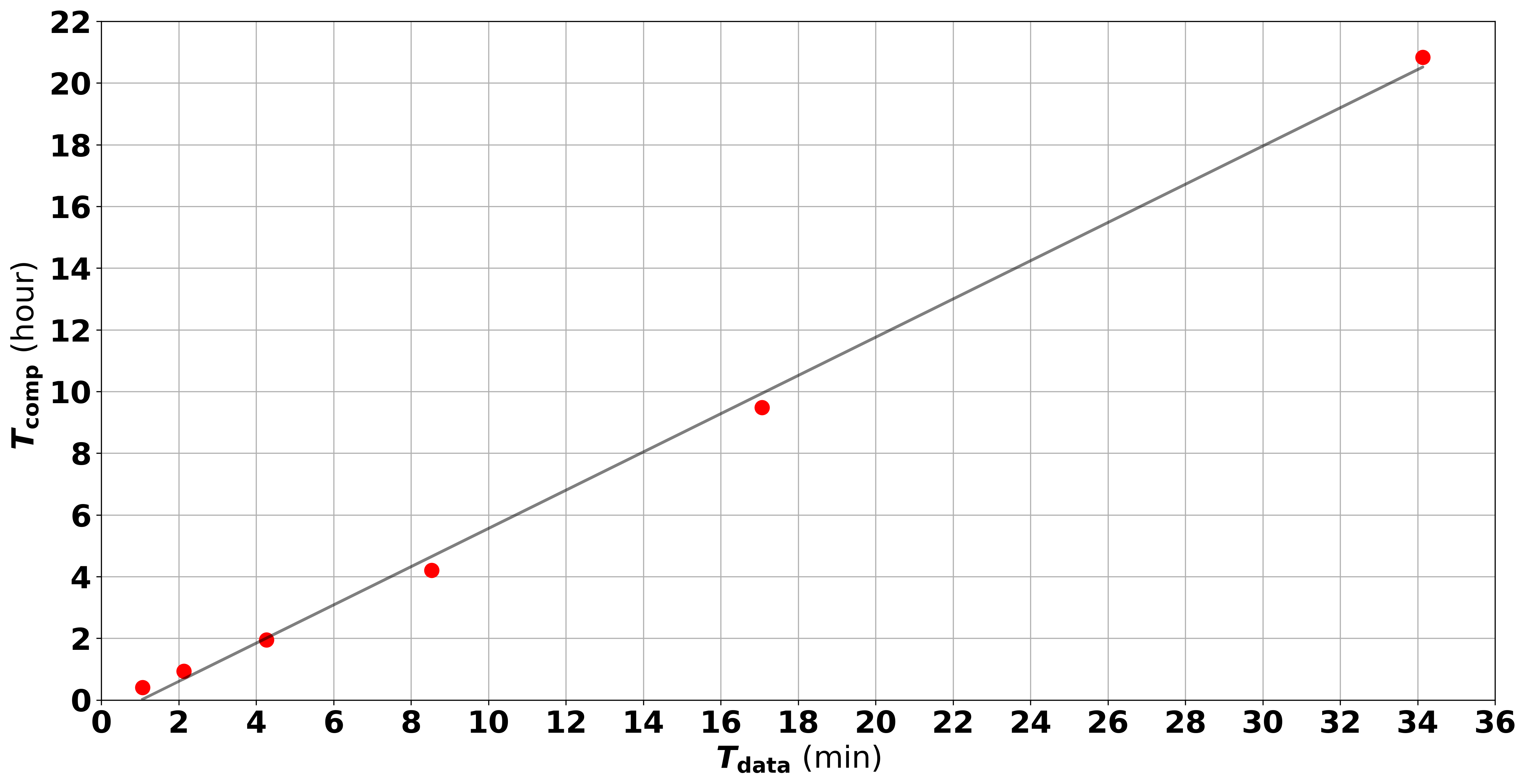}
\caption{\label{fig:computation_time_vs_data_duration}
Shown is the computation time, $T_{\rm comp}$, versus data length, $T_{\rm data}$, for $\niter=1000$.
A linear fit gives 
$37.2$~min of computation time per minute of data. The scatter in the 
computation time across the different PSO runs is in the range of seconds for
all the data lengths above 
and negligible compared to the average.
}
\end{figure}

Our current parallelization strategy results in the time taken by the whole code running across all the nodes and threads of
the cluster being essentially equivalent to the wall-clock computation time, $T_{\rm comp}$, for one thread. 
Fig.~\ref{fig:computation_time_vs_data_duration} shows  $T_{\rm comp}$
as a function of the
duration, $T_{\rm data}$, of the data segment analyzed. 
 
We obtain the $T_{\rm comp}$ values shown by averaging over $4$ PSO runs with $\niter = 1000$ in each run. A linear fit to the data shows that $T_{\rm comp}$ increases at the rate of
$37.2$~min per min of data. 

In the current implementation,  
all operations involved in a single 
fitness evaluation are serialized. However, the principal
steps in a single fitness evaluation are themselves highly parallelizable.
Each fitness evaluation 
involves (i) generating the 
$h_{+,\times}$ waveforms, and (ii) computing the inner product in Eq.~\ref{eq.dot_product}.

Generating the $h_{+,\times}$ waveforms
involves independent calculations of the 
stationary phase approximant of the 2PN waveform~\cite{Blanchet_95}
at each of the Fourier frequencies in the
data. In addition, the inner product in Eq.~\ref{eq.dot_product}
 involves computing the integrand at each Fourier frequency
 independently. 

Thus, an alternative parallelization scheme in a single PSO run 
is to invert the current scheme by
evaluating the fitness values of the $N_p$
particles serially and parallelizing over
the array operations involved in each fitness evaluation. If the computation time for a single 
fitness evaluation is $T_{\rm ff}$, the current $T_{\rm comp}$ is 
approximately $N_{\rm iter}T_{\rm ff}$. With the alternative scheme 
$T_{\rm comp} \approx N_pN_{\rm iter}(T_{\rm ff}/N_{\rm threads})$,
where $N_{\rm threads}$ is the number of threads used in parallelizing the 
fitness function evaluation.
Thus, $N_{\rm threads}$ must be substantially greater than
$N_p$ in order to make the switch to the alternative scheme 
beneficial. 

With high performance 
computing moving increasingly towards massive hardware-level parallelism,
such as the {\sc Intel Xeon Phi Knights Landing} (KNL) processor with $272$
threads or  Graphics Processing Units (GPUs) such as the {\sc Nvidia V100} with
$5120$ threads, we are already in the regime where $N_{\rm threads} \gg N_p$.
Thus, the alternative scheme described above could lead to a substantially 
faster implementation of the fully-coherent all-sky search.

 At present, the only objective  metric for 
comparison of computational costs between the PSO-based search and the
fully-coherent all-sky search pipeline, {\sc LALInference}~\cite{veitch2015parameter}, used for the analysis of LIGO-Virgo
data is the total number of fitness (i.e., likelihood) evaluations. 
The number of fitness evaluations reported in~\cite{veitch2015parameter} range
between $10^6$ and $10^8$ for the Markov Chain Monte Carlo (MCMC) algorithm
using a network of first generation detectors and a data segment length of $32$~sec. The PSO-based search for second generation detectors continues to
maintain the same number of fitness evaluations as found in WM for
a first generation network, ranging from
a maximum of $2.4\times 10^5$ to $4.8\times 10^5$ for the
$64$~sec and $30$~min long data segments
respectively. The number of fitness evaluations per independent run of PSO is a factor of $4$ and $8$ smaller for the respective data lengths.

While the PSO-based search incurs a drastically smaller number of fitness evaluations to compute the coherent search statistic and point estimate of the signal parameters, it should be emphasized that MCMC also provides information about the shape of the posterior probability density function. It may be possible to use the PSO-based search results as a seed to focus the MCMC sampling to a smaller region of parameter space and reduce the computational cost of the latter.

It is also worth noting here that,
 compared to an MCMC search, the PSO-based search 
involves very few tunable parameters,  namely, only $\nruns$ and $\niter$. 

\section{Conclusion}
\label{Conclusion}
A fully-coherent all-sky search for CBC signals  
is implemented that uses PSO as the global optimizer of the joint
likelihood function of data from a network of second generation detectors.
The results presented here show that  
PSO allows  a significant reduction in the computational cost of 
this search with negligible loss in sensitivity.  
 
At a FAR of 1~event/year, the PSO-based search achieves
a detection probability of  $\approx 0.7$ at ${\rm SNR}_{\rm opt} = 9.0$
for a $20.8$~sec long signal, corresponding to an equal mass binary with $14.5$~$M_\odot$ components. For the same signal, 
the detection probability is $\approx 0.99$
at ${\rm SNR}_{\rm opt} = 11$.
These results
demonstrate that the reduced computational cost is achieved without 
any significant loss of sensitivity in an astrophysically relevant range
of signal strength and binary masses. 

Low mass binaries ($O(1.4)$~$M_\odot$ components) pose an extreme challenge
to fully-coherent all-sky searches with second generation detectors because
the 
signal length becomes $O(30)$~min. We tested the PSO-based search 
on a $23$~min signal, corresponding to an equal mass binary with
$1.506$~$M_\odot$, embedded in $30$~min long data segments.
We find that the PSO-based search
continues to perform 
well  with $\leq 6\%$ loss in
detection probability at ${\rm SNR}_{\rm opt}=11$ 
for a realistic detection threshold of $\approx 10$
on the coherent search statistic.

Going beyond Fisher information analyses and Bayesian credible regions obtained with single data realizations, 
the reduced computational cost afforded by PSO 
allows Frequentist parameter 
estimation errors using multiple data realizations to be obtained. 
A limited study in this paper 
shows several interesting issues such as (i) 
variations in localization errors
caused by the condition number of the network
antenna pattern matrix, (ii) non-Gaussianity in the sky localization estimation error, and (iii)
degenerate locations on the sky.  
A more extensive and realistic parameter estimation
study using the PSO-based search will be
carried out in the future.

The prospects for further reductions in the computation time of the 
PSO-based search are very promising. As discussed in the paper, 
existing and upcoming hardware-level support for 
massive parallelism could make an alternative parallelization scheme 
significantly faster than the current one. Incorporating
faster waveform generation schemes, such as the one proposed in~\cite{0264-9381-34-11-115006},
could reduce the computation time even further. It may even be 
possible to squeeze out a factor of few from the current code with 
more efficient implementations of the mathematical formalism.
We are currently  exploring these possibilities and hope to further
push the fully-coherent all-sky search towards real time processing of data.

\section*{Acknowledgements}
The contribution of S.D.M. to this paper was supported by National Science Foundation (NSF)
grants PHY-1505861 and HRD-0734800. 
We acknowledge the Texas Advanced Computing Center (TACC) at The University of Texas at Austin for providing HPC resources that have contributed to the research results reported within this paper.
URL: http://www.tacc.utexas.edu
Some of the pilot studies were carried out on ``Thumper", a computer cluster funded by NSF MRI grant CNS-1040430 at The University of Texas Rio Grande Valley.

\bibliography{references.bib,refs_soumya_dod.bib,pta4gw.bib,sdmLasso.bib,mohanty_bib.bib}

\end{document}